\documentstyle[12pt,epsf]{article}
\begin{document}
\title{Finite energy/action solutions of $p_1$ Yang--Mills equations
on $p_2$ Schwarzschild and deSitter backgrounds in dimensions $d\ge 4$}
\author{{\large Y. Brihaye}$^{\ddagger}$
{\large A. Chakrabarti}$^{\diamond}$
and {\large D. H. Tchrakian}$^{\dagger \star}$ \\ \\
$^{\ddagger}${\small Physique-Math\'ematique, Universit\'e de 
Mons-Hainaut, Mons, Belgium}\\ \\
$^{\diamond}${\small Centre de Physique Th\'eorique, 
Ecole Polytechnique, F-9118 Palaiseau
Cedex, France}\\ \\
$^{\dagger}${\small Department of
Mathematical Physics, National University of Ireland Maynooth,} \\
{\small Maynooth, Ireland} \\ \\
$^{\star}${\small School of Theoretical Physics -- DIAS, 10 Burlington Road,
Dublin 4, Ireland }}

\date{}
\newcommand{\dd}{\mbox{d}}\newcommand{\tr}{\mbox{tr}}
\newcommand{\ee}{\end{equation}}
\newcommand{\be}{\begin{equation}}
\newcommand{\ii}{\mbox{i}}\newcommand{\e}{\mbox{e}}
\newcommand{\pa}{\partial}\newcommand{\Om}{\Omega}
\newcommand{\vep}{\varepsilon}
\newcommand{\bfph}{{\bf \phi}}
\newcommand{\lm}{\lambda}
\def\theequation{\arabic{equation}}
\renewcommand{\thefootnote}{\fnsymbol{footnote}}
\newcommand{\re}[1]{(\ref{#1})}
\newcommand{\R}{{\rm I \hspace{-0.52ex} R}}
\newcommand{\N}{{\sf N\hspace*{-1.0ex}\rule{0.15ex}%
{1.3ex}\hspace*{1.0ex}}}
\newcommand{\Q}{{\sf Q\hspace*{-1.1ex}\rule{0.15ex}%
{1.5ex}\hspace*{1.1ex}}}
\newcommand{\C}{{\sf C\hspace*{-0.9ex}\rule{0.15ex}%
{1.3ex}\hspace*{0.9ex}}}
\newcommand{\eins}{1\hspace{-0.56ex}{\rm I}}
\renewcommand{\thefootnote}{\arabic{footnote}}

\maketitle
\begin{abstract}
Physically relevant gauge and gravitational theories can be seen as special members of hierarchies of more
elaborate systems. The Yang-Mills (YM) system is the first member of a hierarchy of Lagrangians which we
will index by $p_1$, and the Einstein-Hilbert (EH) system of general relativity is the first member of
another hierarchy which we index by $p_2$. In this paper, we study  the classical equations of the
$p_1 = 1,2$ YM hierarchy considered in the background of special geometries (Schwarzschild, deSitter,
anti-deSitter) of the $p_2=1,2,3$ EH hierarchy. Solutions are obtained in various dimensions and lead to
several examples of non-self-dual YM fields. When $p_1=p_2$ self-dual solutions exist in addition. Their
action is equal to the Chern-Pontryagin charge and can be compared with that of the non-self-dual solutions.

\end{abstract}
\medskip
\medskip
\newpage

\section{Introduction}

The main aim of this work is the study of non--selfdual
Yang--Mills (YM) fields in $d\ge 4$ dimensions on fixed gravitational backgrounds
in $d$-dimensions, extending the work of \cite{BCC1,BCC2,C} in $4$-dimensions. As in \cite{BCC1,BCC2,C},
we restrict to Schwarzschild and deSitter metrics. By YM fields here we mean the solutions
to the hierarchy of $p$-YM systems~\cite{T}, whose $p=1$ member is the usual YM system,
and the generic $p$-th member involves the curvature $2p$-form in lieu of the usual YM $2$-form
curvature.

Here we present nonselfdual solutions in curved backgrounds. Such
solutions in flat spaces would be particularly interesting. Even the
complex one in dimension $4$, which was obtained~\cite{BCC2,C} exploiting
the conformal flatness of the deSitter metric in this case, is worthwhile.
(See the remarks and the references in \cite{BCC2} concerning the
relevance of complex saddle points.)

We have a definite reason for considering the hierarchy of YM systems rather than restricting to the
usual ($p=1$) member only, other than the fact that it is quite natural to do so in dimensions $d>4$.
As highlighted in the work of Refs.~\cite{BCC1,BCC2,C}, in $d=4$ dimensions and with Euclidean signature,
the selfdual ($p=1$) YM field can be constructed from the double--selfdual gravitational field by
constructing the $SU(2)$ YM connection from the corresponding gravitational spin-connection~\cite{CD}.
In this case, it is known that the gravitational field equations are automatically satisfied, and so are
the YM equations by virtue of the selfduality\cite{ft1}. Clearly then, the YM action will be equal to the
Chern--Pontryagin (C-P) charge. Thus, if one finds non--selfdual solutions to the
YM equations in the said double--seldual gravitational background, the value of the action will differ
from that of the self--dual YM fields, and it is interesting to compare it with the value of the C-P
charge originating from the double--selfdual metric of the previous case~\cite{BCC1,CD}. For the usual
($p=1$) case a selfdual YM field can thus be related to a double--selfdual gravitational metric
only for $d=4$, and it is our desire to carry it
out in $d>4$, as a subsidiary motive of our investigations, that leads us to consider the hierarchy
of YM systems.

The main interest of the present work remains the investigation of non--selfdual YM fields.
In flat $4$ dimensional Euclidean space no explicit real nonselfdual solutions are known, {\it cf.} the
explicit complex non--selfdual solutions in~\cite{C} for gauge group $SU(2)$. (Implementing instantons and antiinstantons in commuting subgroups of
$SU(N)$ for $N$ sufficiently large one can evidently obtain nonselfdual solutions, but here we consider specifically $SU(2)$.) For $SU(2)$
only selfdual solutions
are known, the value of the YM action pertaining to these being equal to the
Chern--Pontryagin (C-P) charge. It was found in
\cite{BCC1} however that on fixed Schwarzschild and deSitter backgrounds, there were
nonselfdual solutions, the values of whose actions differed from that of
the C-P charge. In particular, in the case of
Schwarzschild background, the value of the nonselfdual action, which turned out to be real,
was slightly smaller than that of the selfdual action. Thus, the study of YM fields on fixed backgrounds
is of general interest in the context of nonselfdual fields, and in the special case(s) where
selfdual solutions on the same background also exist, then it is of particular interest to
compare the value of the nonselfdual action with that of the selfdual one. The reason for
considering the hierarchy of YM models is precisely because the $p$-th member of these, on $4p$ dimensions,
does support selfdual solutions. Indeed, the situation in $4$ dimensions, where the YM field
constructed from the double--selfdual spin--connectionon is
automatically (single) selfdual, occurs in all $4p$
dimensions for the $p$-th member of the YM hierarchy on a background whose $2p$-form Riemann tensor
is double--selfdual. We shall refer these $4p$-dimensional gravitational systems as the hierarchy of
Einstein--Hilbert (EH) systems, or, generalised EH systems~\cite{CT2}.

The hierarchy of (gravitational) EH systems in all even dimensions was previously studied in
Refs.~\cite{CT1} and \cite{OT}, and more recently in \cite{CT2}. In \cite{CT1} it was shown that
in $4p$ dimensions, the $2p$-form Riemann tensors of the deSitter and Fubini-Study
metrics were double--selfdual, and in \cite{CT2} this was done for Schwarzschild like metrics.
In \cite{OT} it was shown that
the YM connections constructed from the corresponding spin-connections~\cite{CD} in all even dimensions
yielded $2N$-form YM curvatures which satisfied the (single--)selfduality conditions of the YM
hierarchy~\cite{T}. This construction is possible in all even dimensions only when the curved space is
a compact coset space, but in general can also be carried out if we restrict the dimensions to $4p$. In
the present work, where YM fields (both selfdual and nonselfdual ) on Schwarzschild backgrounds
are considered, we restrict to $4p$ dimensions, and we label the EH systems also with the label $p$.

Since we will consider the $p$-th members of both YM and EH hierarchies in all $d$ dimensions, with
$d\neq 4p$ for the non--selfdual cases, we will henceforth label the members of the YM hierarchy with
$p_1$ and the members of the EH hierarchy with $p_2$. For a discussion of the YM and EH hierarchies, we
refer to \cite{T} and \cite{OT}, but here we simply give the definitions for the Lagrangians of these
systems for the particuar examples that will be employed in the present work.

Thus, concerning the YM systems, the $p_1 =1$ and the $p_1 =2$ systems read respectively
\begin{eqnarray}
{\cal S}_{(1)} &=& \mbox{Tr} \: F^{\mu \nu}F_{\mu \nu}
, \label{ip=1} \\
{\cal S}_{(2)} &=& \mbox{Tr} \: F^{\mu \nu \rho \sigma}F_{\mu \nu \rho \sigma}
\: .\label{ip=2}
\end{eqnarray}
with the following  definitions for the field strenghts (the potentials
are antihermitian, the gauge group will be specified later)
\be
\label{fdef}
F_{\mu\nu}=\partial_{\mu}A_{\nu}-\partial_{\nu}A_{\mu}+[A_{\mu},A_{\nu}] \ \ ,
\ee
\be
\label{2.16}
F_{\mu \nu \rho \sigma}=\{ F_{\mu [\nu},F_{\rho \sigma]} \}\: 
\ee
and the square bracket on the indices $[\nu \rho \sigma]$ implies cyclic symmetry.
The final normalisations of both \re{ip=1} and \re{ip=2} will be fixed such that the value of the action
pertaining to the spherically symmetric self--dual solutions on Schwarzschild background (with Euclidean
signature), is set equal to one.

Concerning the EH systems, these are the $p_2 =1$, $p_2 =2$ and the $p_2 =3$ systems
\begin{eqnarray}
{\cal L}_{(1)} &=& \vep^{\mu_1 \mu_2 \nu_1 ..\nu_{d-2}}e_{\nu_1}^{n_1}..e_{\nu_{d-2}}^{n_{d-2}}
\vep^{m_1 m_2 n_1 ..n_{d-2}}\: R_{\mu_1 \mu_2}^{m_1 m_2}\: , \label{iip=1} \\
{\cal L}_{(2)} &=& \vep^{\mu_1 ..\mu_4 \nu_1 ..\nu_{d-4}}e_{\nu_1}^{n_1}..e_{\nu_{d-4}}^{n_{d-4}}
\vep^{m_1 ..\mu_4 n_1 ..n_{d-2}}\: R_{\mu_1 \mu_2}^{m_1 m_2} R_{\mu_3 \mu_3}^{m_3 m_4}\: ,
\label{iip=2} \\
{\cal L}_{(3)} &=& \vep^{\mu_1 \mu_2 ..\mu_6 \nu_1 ..\nu_{d-6}}e_{\nu_1}^{n_1}..
e_{\nu_{d-6}}^{n_{d-6}}\vep^{m_1 ..m_6 n_1 ..n_{d-2}}\:
R_{\mu_1 \mu_2}^{m_1 m_2} R_{\mu_3 \mu_3}^{m_3 m_4} R_{\mu_5 \mu_6}^{m_5 m_6}\: , \label{iip=3}
\end{eqnarray}
in $d$ dimensions. The corresponding (generalised) Einstein equations are given in \cite{CT1,OT,CT2}.

In the self--dual cases, with $p_1 =p_2 =d/4$, the stress tensor due to the YM fields vanishes identically
so that the (generalised) Ricci scalars \re{iip=1}-\re{iip=3} vanish for these field configurations.
Thus the action of the YM fields equals the Chern-Pontryagin charge. To calculate the action integral in
(static) Schwarzschild backgrounds in such cases, one has to integrate the time variable over one
period. This is the period associated with the desingularisation of the Schwarzschild metric by
introducing Kruskal like coordinates in the case of Euclidean signature. For the arbitrary dimensional
case, and for the dynamics determined by the $p_2$ EH system, this period was calculated~\cite{CT2} to be
\be
\label{i1}
P_{(p_2)}=\frac{4\pi {\cal K}p_2}{d-2p_2 -1}\: ,
\ee 
where ${\cal K}$ is a parameter\cite{ft2} in the (hierarchy of)
Schwarzschild metric(s)given in~\cite{CT2}.
In our calculations of the actions below we will suppress the factor \re{i1} contributed by the (Euclidean)
time integration, since we are only interested in the relative values of the selfdual and non--selfdual actions
for any given EH--YM system.

In \cite{BCC1}, very simple nonselfdual solutions for the $d=4$ (usual)
$p_1=1$ YM fields in Schwarzschild and deSitter backgrounds (both of the $p_2=1$ member of the EH
hierarchy) in four dimensions were presented. They were further discussed in Refs.~\cite{BCC2}
and \cite{C}. Here they are generalized to all dimensions $d\geq 4$, with Lorentz or Euclidean signature
and $(d-1)$ spatial dimensions, for both $p_1=1$ and $p_1=2$ members of the YM hierarchy, on the
backgrounds of the $p_2=1,2,3$ members of the EH hierarchy. The corresponding
constructions for members of the $p_1\ge 3$ on various $p_2$ backgrounds can be given systematically.
As in the $4$-dimensional $p_1=1$ YM case~\cite{BCC1} on $p_2=1$ Schwarzschild and deSitter backgrounds,
where these solutions are found in closed form, it turns out that in the $d$-dimensional ($d>4$) $p_1=1$
YM case on these $p_2=1$ backgrounds, non--selfdual solutions can also be constructed analytically
in closed form using the same procedure. On $p_2 \ge 2$ backgrounds however, non--selfdual solutions of
the $p_1=1$ YM system could only be eveluated by numerical construction.
In the $p_1=2$ (as well as all $p_1>2$) YM case(s) likewise, the non--selfdual solutions on $p_2$ EH
background could be constructed only using numerical integrations irrespective of $p_2$. These are the
main results of the present work and are presented in Section {\bf 2}. The corresponding constructions
for members of the $p\ge 3$ can be given systematically.

For the $p_1=1$ and $p_1=2$ systems in $d=4$ and $d=8$ dimensions respectively (with Euclidean signature),
the selfdual solutions are considered in detail for $p_2=1$ and $p_2=2$ Schwarzschild metrics,
respectively. These results are needed for the comparison of the Euclidean actions of the selfdual
and nonselfdual solutions, which we carry out for these cases. The corresponding analysis with
deSitter backgrounds is not carried out in detail since in that case there are no (real) non--selfdual
solutions. Again, extension to the generic ($p_1=p_2$)-th selfdual cases can be carried out systematically.
These results are presented in Section {\bf 3}. In Section {\bf 4}, we give
a short summary and discussion.

Solutions on the anti-deSitter backgrounds of $p_2=p$ EH, to the $p_1=p$ YM
in $d=4p$ dimensions, are studied in the Appendix.

\section{Non--selfdual solutions on fixed backgrounds}

This Section is subdivided in three Subsections. 
In the first one, we state our Ansatz for the YM
fields on fixed Schwarzschild and deSitter curved 
backgrounds using the Kerr-Schild parametrisation
of these metrics and give the Euler--Lagrange equations whose non--selfdual solutions we seek.
In the second and third Subsections, we present the non--selfdual solutions on
Schwarzschild and deSitter backgrounds, respectively.

Since all the actual calculations involved in this, and the next Section, involve only Euler--Lagrange
equations and selfduality conditions of members of the YM hirerchy, we will not be engaged in a description
of the hierarchy of Einstein--Hilbert (EH) systems. The only information we need here in this respect are
the actual functions parametrising the Schwarzschild and deSitter metrics pertaining to these hierarchies,
which we have stated when needed below in \re{2.8} and \re{2.9}. For details of their derivations, we refer to
\cite{CT2}. Similarly for the hierarchy of YM systems we simply state their field equations
\re{2.15a}-\re{2.15b} and, refer to \cite{T} for their general details and to
\cite{OT} for their properties relative to the EH hierarchy.

\subsection{Ans\"atze and YM equations}

There are several ingredients needed for making the spherically symmetric Ansatz. 
One is the definition of spin
matrices, or the gamma matrix representations of the generators of $SO(d)$ 
(the choice of this gauge group
is dictated by our requirement of spherical symmetry in $d$ and $d-1$ dimensions).
Another ingredient is the
parametrisation of the components of the metric for the (fixed) curved space 
on which the YM fields are
defined. The ansatz for the gauge potentials has then to be
specified and finally the equations derived.
We deal with these items in that order below.

\subsubsection{Gauge group and representation}

Even though our dominant motivation here is the study of non--selfdual solutions, let us start with the case
of selfdual solutions which we shall consider in the next Section for comparison with the results of this
Section. This puts a restriction on the 
$2^{d/2}\times 2^{d/2}$-dimensional representations of the $SO(d)$ matrices in
spacetime with $d$-dimension. 
The representations of the gauge groups of the YM fields pertaining to the
$p$-th member of the YM hierarchy are chosen such that in $4p$ dimensions, there exist selfdual solutions
on $\R^{4p}$~\cite{T}, namely that the gauge group is represented by
$2^{(d-2)/2}\times 2^{(d-2)/2}=2^{2p-1}\times 2^{2p-1}$ (left or right) chiral $SO(4p)$ matrices,
denoted by $SO_{\pm}(4p_1)$. This choice of gauge group representation also makes it possible to construct
single--selfdual $p_1=p$ --YM fields on $p_2=p$ --EH backgrounds \cite{OT}. Examples of the latter are
known~\cite{CT1,OT,CT2}, and of these the deSitter~\cite{CT1} and Schwarzschild~\cite{CT2} will concern us
in this work.

The representations of the algebra of $SO_{\pm}(4p)$ employed in our Ans\"atze are denoted by
\be
\label{2.1}
\Sigma_{ab}=-\frac{1}{4}\Sigma_{[a}\tilde \Sigma_{b]}=-\frac{1}{8}(\eins 
\pm \Gamma_{4p+1})[\Gamma_a ,\Gamma_b]
\: , \quad (a,b=1,2,..,d)
\ee
where the square brackets on the $4p$ component indices $[ab]$ imply antisymmetrisation, $\Gamma_a$
denote the $d=4p$ --dimensional Gamma matrices and $\Gamma_{4p+1}$ is the corresponding chiral matrix.

As we shall be concerned with $p=2$-YM systems in detail in this and the next Section,
it is convenient to define the totally antisymmetric $2p$-form tensor--spinor matrix
\be
\label{2.2}
\Sigma_{abcd} =\{ \Sigma_{a[b} ,\Sigma_{cd]} \}
\ee
using  the same notation as in (\ref{2.16}).
In addition, we state an identity
which will be useful in the next Section,
\be
\label{2.3}
\{ \Sigma_{ab} ,\Sigma_{cd}\}=\frac{1}{3}\Sigma_{abcd}
-\frac{1}{2}(\delta_{ac}\delta_{bd}-\delta_{ad}\delta_{bc})\eins \: .
\ee

Let us now relax the constraints necessitated by the requirement of YM systems to support selfdual fields,
thus allowing $d$ dimensional $SO(d)$ systems, without requiring that $d=4p$. For even $d$
this could allow the assignment of chirally symmetric $2^{d/2}\times 2^{d/2}$ representations for the spin
matrices. These are not the representations we assign to the even $d$ dimensional gauge algebras,
but restrict to the chirally asymmetric (left/right) $2^{(d-2)/2}\times 2^{(d-2)/2}$ spin matrices
defined by \re{2.1}. The reason is that these are the relevant representations for the selfdual cases, which
we wish to compare eventually against the non--selfdual cases under consideration. For the static field
configurations (deSitter or Schwarzschild) we will be concerned with, the spherical symmetry will be imposed
in the odd, $d-1$, dimensions, when there there exits 
no chiral matrix. The representations of the spin
matrices in this case are defined by the 
$2^{(d-2)/2}\times 2^{(d-2)/2}$-dimensional  $SO(d-1)$ matrices $\Sigma_{ij}$,
with $(i,j=1,2,..,d-1)$ given by \re{2.1}, or equivalently by
\be
\label{2.1a}
\Sigma_{ij}=-\frac{1}{4}[\Gamma_i ,\Gamma_j]
\ee
in terms of the $(d-2)$
Gamma matrices $\Gamma_1 ,\Gamma_2 ,
..,\Gamma_{d-2}$ (with dimension $2^{(d-2)/2}\times 2^{(d-2)/2}$), 
supplemented by their chiral matrix $\Gamma_{d-1}$. In what follows, the precise
dimensionality of the representations of $\Sigma_{ij}$ will 
not matter except in the (canonical) dimensions
where selfdual YM fields are supported, inosfar as the value of the 
action densities depend on these.
Otherwise their only important feature will be the fact that they
satisfy the algebra of $SO(d)$ or $SO(d-1)$ respectively.

\subsubsection{The metric}

Next, we give the Kerr-Schild parametrisation of the background metric, which will be employed in the
Ansatz. We consider spherically symmetric, static metrics for dimensions $d\geq 4$
\be
\label{2.4}
g_{\mu\nu}=\eta_{\mu\nu}+l_{\mu} l_{\nu}\: , \quad g^{\mu\nu}=\eta^ {\mu\nu}-l^{\mu} l^{\nu}
\ee
where
\[
\eta_{00} = \eta^{00} = -1\: , \quad \eta_{ij} =\eta^{ij} = \delta_{ij}
\]
and
\be
\label{2.6}
l^{\mu}l^{\nu}\eta_{\mu\nu} = l^{\mu}l^{\nu}g_{\mu\nu} =0\: .
\ee

For static spherical symmetry $l_0$ is a function of $r$ only, where
$$r^2= \sum_{i=1}^{d-1} x_i^2$$
and
$$l_i=l_0 \frac{x_i}{r}=l_0 \hat x_i \: ,  \qquad (i=1,2, \dots,d-1)$$
satisfying
\be
\label{2.7}
-l_0^2 +\sum_i  l_i^2 =0
\ee

The Schwarzschild metric in $d$ dimensions pertaining to the usual ($p_2=1$) EH system, was given by
Myers and Perry~\cite{MP}. The correspoding result in $d$ dimensions pertaining to the generic $p$ EH
system was given recently in Ref.~\cite{CT2}. For the generic case we have
\be
\label{2.8}
l_0^2 = Cr^{-\left(\frac{d-2p-1}{p}\right)}\: , \quad (C>0)
\ee
and for deSitter metric, for all $d$ pertaining to all $p$ EH systems,
\be
\label{2.9}
l_0^2 = \Lambda r^2 \qquad (\Lambda >0)\: .
\ee

The standard form in spherical coordinates is given by
\be
\label{2.10}
ds^2 = - N dt^2 + N^{-1} dr^2 + r^2 d\Omega_{(d-2)} ,
\ee
where $ d\Omega_{(d-2)}$ is the line element on the unit $(d-2)$-sphere and
\be
\label{2.11}
N=( 1-l_0^2 )
\ee
The coordinate transformation relating \re{2.4} and \re{2.10}, namely
\begin{eqnarray}
x_0 = t +\int {\frac{dr}{N}} - r
\end{eqnarray}
does not affect our particularly simple ansatz for the gauge potentials to follow (with
$A_t = A_r =0$).

After constructing the solutions using \re{2.4} the passage to Euclidean signature is
best considered ( rather than introducing imaginary $l_0$) by directly starting from
\re{2.10}, leading to 
\begin{eqnarray}
\label{metric}
ds^2 =  N dt^2 + N^{-1} dr^2 + r^2 d\Omega_{(d-2)} 
\end{eqnarray}

\subsubsection{The gauge potentials}

The ansatz for the gauge potentials is
\begin{eqnarray}
A_0 &=&0 \label{2.11a}  \\
A_i &=& r^{-1} \: \Bigl(K(r)- 1\Bigr)\: \Sigma_{ij}\hat x_j  \quad (i=1,2, \dots,d-1) \label{2.11b}
\end{eqnarray}
with $\Sigma_{ij}$ defined according to \re{2.1}, or equivalently \re{2.1a}, 
to be the spinor representations of $SO(d-1)$. 
The corresponding YM curvature (\ref{fdef})
has then the following components
\begin{eqnarray}
F_{0i} &=& 0 \label{2.12a} \\
F_{ij} &=& V_1 \: \Sigma_{ij} +V_2 \: \hat x_{[i}\: \Sigma_{j]k} \: \hat x_k \label{2.12b}
\end{eqnarray}
where as before, the square brackets on the subscripts $[ij]$ imply antisymmetry, and,
\begin{eqnarray}
V_1 &=& -r^{-2}(K^2 -1) \label{2.13a} \\
V_2 &=& -r^{-2}[rK'-(K^2 -1)] \label{2.13b} \: .
\end{eqnarray}

Now using \re{2.4} we have
\begin{eqnarray}
F^{0i} &=& g^{0\mu}g^{i\nu}F_{\mu\nu} \nonumber \\
       &=& l_0^2 K'\: \Sigma_{ik}\hat x_k \label{2.14a} \\
F^{ij} &=& g^{i\mu}g^{j\nu}F_{\mu\nu} \nonumber \\
& = & W_1 \: \Sigma_{ij} +W_2 \: \hat x_{[i}\: \Sigma_{j]k} \: \hat x_k \label{2.14b}\: ,
\end{eqnarray}
with
\begin{eqnarray}
W_1 &=& -r^{-2}(K^2 -1) \label{2.13c} \\
W_2 &=& -r^{-2}[rNK'-(K^2 -1)] \label{2.13d} \: .
\end{eqnarray}

\subsubsection{The equations}

The Euler-Lagrange equations of motion are expressed very simply since
$\mid g\mid =1$ for the KS metric. 
For the $p=1$ and $p=2$ YM systems they are, respectively
\begin{eqnarray}
D_{\mu}\: F^{\mu \nu} &=& 0 \label{2.15a} \\
\{ F_{\rho \sigma}\: ,\: D_{\mu}\: F^{\mu \nu \rho \sigma}\} &=& 0 \label{2.15b}\: .
\end{eqnarray}
with the definitions \re{fdef},\re{2.16}.

After some straightforward but laborious calculations, substituting \re{2.11a}, \re{2.11b}, and \re{2.14a},
\re{2.14b}, into \re{2.15a} and \re{2.15b}, we find first that the Gauss law equations
\[
D_{\mu}\: F^{\mu 0} = 0\: \quad {\rm and} \quad \{ F_{\rho \sigma}\: ,\: D_{\mu}\: F^{\mu 0 \rho \sigma}\} = 0
\]
are identically satisfied, and the remaining components of \re{2.15a} and \re{2.15b} yield simply
\begin{eqnarray}
D_{\mu}\: F^{\mu j} &=& r^{-(d-3)}\left( [Nr^{d-4}K']'-(d-3)r^{d-6}K(K^2 -1) \right) \times
\Sigma_{jk}\: \hat x_k \label{2.17a} \\
\{ F_{\rho \sigma}\: ,\: D_{\mu}\: F^{\mu j \rho \sigma}\} &=& r^{-(d-5)}V_1 \left( [Nr^{d-8}(K^2 -1)K']'
-(d-5)r^{d-10}K(K^2 -1)^2 \right) \: \times \nonumber \\
&&\qquad \qquad \qquad \qquad \qquad \times \: 3(d-3)(d-4)\: \Sigma_{jk}\: \hat x_k \label{2.17b}\: .
\end{eqnarray}

The Euler-Lagrange equations of the one dimensional subsystems of the $d$ dimensional $p=1$ and $p=2$ YM
systems in the static spherically symmetric background specified by \re{2.4} can readily be read off
\re{2.17a} and \re{2.17b}, as single equations for the functions $K(r)$. Indeed, it is easy to read off the
equations for the arbitrary $p$ YM systems:
\be
\label{2.18}
[Nr^{d-4p}(K^2 -1)^{p-1}K']'=[d-(2p+1)]r^{d-2(2p+1)}K(K^2 -1)^p \: .
\ee

We discussed in the Introduction, the Einstein--Hilbert (EH) hierarchy in parallel with the YM hierarchy.
The former determines the metric on whose background the YM fields are studied. In the context of \re{2.18},
it is the function $N(r)$ that is determined by the dynamics of the relevant EH system. Since in this
Section we are interested in solutions on fixed backgrounds, there is no reason to privilege any
particular member of the EH systems for characterising the function $N$ given by \re{2.8}, \re{2.9} and
\re{2.11}. Using the notation introduced in
Section {\bf 1}, we label this function by the index $p_2$ pertaining to the EH hierarchy, namely as
$N_{(p_2)}$, and simultaneously rename the index $p$ in \re{2.18} $p_1$. Equations \re{2.18} now are
expressed more specifically as
\be
\label{2.18a}
[N_{(p_2)}r^{d-4p_1}(K^2 -1)^{p_1-1}K']'=[d-(2p_1+1)]r^{d-2(2p_1+1)}K(K^2 -1)^{p_1} \: .
\ee

For use in comparing the actions of non--selfdual and selfdual $p=1$ and $p=2$ YM systems in $d=4$ and
$d=8$ respectively in the next Section, we present the action densities of these two systems for the
fields \re{2.12a}, \re{2.12b}, \re{2.14a} and \re{2.14b}, for which we will give the non--selfdual solutions
in the following two Subsections. Because of the vanishing of $F_{0i}$, \re{2.12a}, these action densities
coincide with the (Euclidean) energy densities. They are, for the $p=1$ YM system,
\begin{eqnarray}
{\cal S}_{(1)} &=& \mbox{Tr} \: F^{\mu \nu}F_{\mu \nu}=\mbox{Tr} \: F^{ij}F_{ij} \label{p=1} \\
&=& 2^{\frac{d-6}{2}}
(d-2)[(d-3)W_1 V_1 +2(W_1 -W_2)(V_1 -V_2)] \nonumber \\
&=& 2^{\frac{d-6}{2}}
\frac{(d-2)}{r^4}\left[2r^2NK'^2 +(d-3)(K^2 -1)^2 \right] \label{2.19}\: ,
\end{eqnarray}
and for the $p=2$ YM system,
\begin{eqnarray}
{\cal S}_{(2)} &=& \mbox{Tr} \: F^{\mu \nu \rho \sigma}F_{\mu \nu \rho \sigma}=\mbox{Tr} \: F^{lijk}F_{lijk}
\label{p=2} \\
&=& 2^{\frac{d-10}{2}}
(d-2)(d-3)(d-4)\times \nonumber \\
&&\times W_1 V_1[(d-5)W_1 V_1 +4(W_1 -W_2)(V_1 -V_2)]
\nonumber \\
&=&2^{\frac{d-10}{2}}(d-2)(d-3)(d-4) \times \nonumber \\
&& \quad \times \frac{(K^2 -1)^2}{r^8} \left[4r^2 NK'^2 +(d-5)(K^2 -1)^2 \right] \label{2.20}\: .
\end{eqnarray}
In \re{2.19} and \re{2.20}, the numerical factors resulting from the traces of the $\Sigma$ matrices are
accounted for. We note here that varying the densities $r^{d-2}\mbox{Tr} \: F^{ij}F_{ij}$ and
$r^{d-2}\mbox{Tr} \: F^{lijk}F_{lijk}$, given by \re{2.19} and \re{2.20} respectively, with respect to $K(r)$,
we obtain the equations \re{2.18a} with $p_1 =1$ and $p_1 =2$ respectively. This is not surprising.

Before proceeding to integrate \re{2.18} for $p=1$ and $p=2$, we make some general remarks. Firstly, equations
\re{2.18} do not satisfy the Painlev\'e criterion~\cite{R}, nonetheless, we find some special solutions.
Perhaps the most remarkable feature of equations \re{2.18}
 is the fact that on flat background,
with $N=1$, there are no solutions so that the only 
non--selfdual solutions are on curved backgrounds.
It is interesting to make some
general observations at this point. We note that in the generic case, in dimensions $d=2p+1$, the right
hand side of \re{2.18} vanishes. In particular, for the $p=1$ this coincides with $d=3$ corresponding to
the Abelian case with only $\Sigma_{12}$. In what follows, we will restrict our attention to $d\geq 4p$ for
each case $p$. Special features arising for $d=4p+1$ will be discussed also.

\subsection{$p_2$ Schwarzschild backgrounds}

We now present the solutions of Eq.\re{2.18a} for the Schwarzschild 
case i.e. for 
\be
\label{nsch}
  N_{(p_2)} = 1 - (\frac{\bar C}{r})^{\frac{d-2p_2-1}{p_2}} \ \ \ , \ \ \
  C \equiv \bar C ^{\frac{d-2p_2-1}{p_2}}
\ee
(where we redefined the parameter $C$ of (\ref {2.8})).
We will limit ourselves to those values of $p_2$ for which the
metric   (\ref{metric}) is asymptotically flat, i.e. to
$p_2 < (d-1)/2$.  Eq.\re{2.18a} has to be solved on the 
interval $[\bar C, \infty]$ with the boundary conditions
\be
    \frac{d-2p_2-1}{p_2}  \bar C K'(\bar C) 
  - (d-2p_2-1) K(\bar C) (K^2(\bar C) - 1) = 0  
\ \ \ , \ \ \ K(\infty) = 1
\ee
which arise by demanding the regularity of the 
solution at $r = \bar C$ and the finitenes of the action.

\subsubsection{$p_1 =1$ , $p_2 =1$ case}

 Solutions to \re{2.18a} with one--node
can be constructed by using the ansatz
\be
\label{2.2.1}
K=\frac{r^{(d-3)}+a\bar C^{(d-3)}}{r^{(d-3)}+b\bar C^{(d-3)}} \ \ .
\ee
This leads to simple algebraic constraints on the parameters $a,b$ 
\begin{eqnarray}
3a+b(2d-7)+(d-1)&=&0 \label{2.2.2a}\\
a(a+b)-(d-5)b&=&0\: . \label{2.2.2b}
\end{eqnarray}
which involves essentially solving only a quadratic equation.
For $d=4$ the old results
\cite{BCC1} are reproduced. 
Of the two real solutions only the one with
\begin{equation}
\label{a,b}
a=-2.366, \quad b=4.098
\end{equation}
(the exact values follow from \re{2.2.2a}-\re{2.2.2b})
gives finite energy (action) for Lorentz (Euclidean) signature.

For $d=5$ one has a very special case, as is evident from \re{2.2.2b}. The solution
\be
\label{2.2.3}
a=0, \quad b= -(4/3)
\ee
leads to a divergent action since the domain of $(r/ \bar C)$ is $[1,\infty]$ and this includes a
zero of the denominator of $K$. But now one can also consider the flat limit as follows.

Setting, for $(d=5)$,
\be
\label{2.2.4}
a=\hat{a}/ \bar C^2, \quad b=\hat{b}/ \bar C^2
\ee
\re{2.2.2a} and \re{2.2.2b} reduce to
\begin{eqnarray}
3(\hat{a}+\hat{b})+4\bar C^2&=&0 \label{2.2.5a} \\
\hat{a}(\hat{a}+\hat{b})&=&0 \label{2.2.5b}\: .
\end{eqnarray}
Hence as $\bar C\rightarrow 0$, setting ($\delta$ being an arbitrary real number)
\be
\label{2.2.6}
-\hat{a}=\hat{b}=\delta^2
\ee
one obtains
\be
\label{2.2.7}
(K-1)=-\frac{2\delta^2}{r^2+\delta^2}\: .
\ee
Substituting \re{2.2.7} in \re{2.12a}, \re{2.12b}, \re{2.14a}and \re{2.14b} it is seen that for the
convention, say,
\[
\epsilon_{1234}=1\: ,\quad  \Sigma_{12}=-\Sigma_{34} \quad (\rm{cyclic})
\]
for the chirally projected $2\times 2 \ \ SO(4)$ generators one obtains the famous BPST
selfdual solution in $d=4$ as the static limit in $d=5$ via our limiting process. Another
convention gives the antiselfdual form.

>From $d=6$ onwards the solutions become complex. The corresponding finite complex
action ,or energy, will be obtained in the following section. The exact values can be
obtained, for any $d$, immediately from \re{2.2.2a}-\re{2.2.2b}. Some numerical values, giving a direct
idea of variation with $d$, are  given below. Both upper or both lower signs are to be
taken. For
\begin{eqnarray}
&&d=6:\quad a=0.500\pm i1.500,\qquad  b=-1.300\mp i0.900 {\nonumber}\\
&&d=7:\quad a=0\pm i1.732, \qquad \qquad b=0.857\mp i0.742 {\nonumber}\\
&&d=8: \quad a=-0.167\pm i1.863,\qquad  b= -0.722\mp i0.621 {\nonumber}\\
&&d=9: \quad a=-0.250\pm i1.984,\qquad  b=-0.659\mp i0.541 {\nonumber}\\
&&d=10: \quad a=-0.300\pm i2.100,\quad b=-0.623\mp i0.485\: . {\nonumber}
\end{eqnarray}

We now come back to the case $d=4$ for which we were able to 
further construct numerically a solution  with two--node. 
It has in particular
\be
\label{2.2.10.a}
     K(1) \approx 0.045 \ \ , \ \ 
     K(2.35) \approx 0.0 \ \ , \ \ 
     K(54.) \approx 0.0 
\ee
and the two profiles of $K$ are plotted on Fig.~1.
In this figure, it is seen that the two solutions for $K(r)$ 
 cross the value $K=0$ at 
nearly the same value $r \approx 2.36$ (although numerically different) 
and that the two-node solution
reaches its asymptotic value $K=1$ very slowly.

The occurence of such a couple of solutions suggests that the action
functional (see Eq. \re{2.2.9} below)
admits an infinite series of extrema
indexed by the number of nodes of the function $K(r)$. 
This is reminiscent to the series of Bartnik-McKinnon~\cite{BM} 
solutions in Einstein-Yang-Mills theory. 
Similar series of particle--like solutions have also been discovered to 
the Einstein-Yang-Mills-Higgs equations \cite{bfm}. For a review, see
\cite{volkov}.

The construction of the (eventual) 
additional solutions in our case would demand 
more numerical analysis and  lies
outside the scope of the present work.

The different solutions can be characterized by their action.
Up to trivial factors this is determined by the following
radial integral
\be
\label{2.2.9}
I_d(p=1) = \int_1^{\infty} dx\: x^{d-6}[2r^2N_{(1)}
(K')^2+(d-3)(K^2-1)^2] \ \ .
\label{ip1}
\ee
For the two solutions of the case $d=4$, we find
\be
\label{2.2.10a}
          I_4^{(1)} \approx 0.959 \ \ \ , \ \ \ I_4^{(2)}\approx 0.992
\ee
respectively or the one--node and two--node solutions.

No solutions other could be constructed numerically for the 
equation
 \re{2.18a} with $p=1$ and  $d > 4$. 
This is probably because, if any, they are not real.

\subsubsection{$p_1 =2$,  $p_2 =2,3,4$ case}

In the case $p_1=2$, we could not find a self consistent
ansatz like (\ref{2.2.1}); however, we manage to construct
a few real solutions numerically.
In analogy with the case $p_1 =1$  above, where the only
real solution was in $d=4p_1 =4$ dimensions, here we expect 
to find real solutions in $d=4p_1 =8$
dimensions. This indeed turns out to be the case. 
In addition, it is possible to
employ the three members labeled by $p_2 =1,2,3$ of the 
EH hierarchy to specify the fixed background.

The non trivial factor of the action integral reads in this case 
\be
\label{2.2.10}
I_d(p_1=2) = \int_1^{\infty} dx\: x^{d-10}[4x^2N_{(p_2)}
(K')^2 (K^2-1)^2 +(d-5)(K^2-1)^4]\: ,
\label{ip2}
\ee

We were able to construct numerically
a solution where $K(r)$ develops one node and another one
where $K(r)$ develops two nodes.
For the three cases $p_2=1,2,3$,
the numerical evaluation of the action/energy
integrals 
\re{2.2.10}  leads to
\begin{eqnarray}
p_2=1 \ \ & \ \ I_8^{(1)} \approx 2.61 \ \ & \ \ I_8^{(2)} \approx 2.95 \label{2.2.12a} \\
p_2=2 \ \ & \ \ I_8^{(1)} \approx 2.38 \ \ & \ \ I_8^{(2)} \approx 2.90 \label{2.2.12b} \\
p_2=3 \ \ & \ \ I_8^{(1)} \approx 2.17 \ \ & \ \ I_8^{(2)} \approx 2.81 \label{2.2.12c}
\end{eqnarray}
Again, the existence of two--node solutions suggests that the
functional \re{2.2.10} admits an infinite series of extrema
indexed by the number of nodes of the function $K(r)$.

Inspection of \re{2.2.12a}, \re{2.2.12b} and \re{2.2.12c} reveals that the action/energy integral $I_{(8)}$
of the $p_1$-th member of the YM hierarchy on the fixed background of the $p_2$-th member of the EH
hierarchy, decreases with increasing $p_2$. This is true for both one--node and two--node
solutions, and we expect it is a general feature for such sytems.

The profiles of the solutions are presented in Fig. 2.
Again in the case $p_2 =1$ we remark that the two solutions
cross $K=0$ for nearly the same value ($r \approx 2.13$) of
the radial variable. This phenomenon is less and less true for increasing $p_2$.

Our numerical analysis of the equations for $d>8$ has lead
to no other solution.

Superpositions (by means of linear combinations)
of the lagrangians with different
values of $p_1$ and/or of $p_2$ could be considered as well, 
leading to a many-parameter  differential equation
but we have not considered such possibilities.

\subsection{deSitter backgrounds}

In this case, there is one single background function $N(r)$ 
for all members $p_2$ of the EH hierarchy,
in all dimensions $d$.
The relevant equation to solve (\ref{2.18a}) with
\be
  N=(1- \Lambda r^2)  
\ee
(see  \re{2.9} and \re{2.11}).
We therefore consider only the different members of the 
YM hierarchy and label them with $p_1 =p$ throughout this section. 

The equation has to be solved on the interval 
$r \in [0,\Lambda^{-1/2}]$
with the boundary conditions
\be
\label{2.3.8}
K(0) = 1 \ \ \ ,\quad
2 \bar \Lambda K'(\bar \Lambda)+
(d-2p-1)K(\bar \Lambda)(K^2(\bar \Lambda)-1) = 0\: .
\ee 
where for brevity we have used $\bar \Lambda \equiv \Lambda^{-1/2}$.  

Again, in practice there is a marked difference between the $p=1$
case and all others. It turns out that for $p=1$, very simple solutions can be found in closed
form, while for all other cases the solutions can be constructed only numerically.

\subsubsection{$p=1$ case}

In this case, solution can be constructed algebraically
by using the ansatz
\be
\label{2.3.1}
K=\frac{1+a\Lambda r^2}{1+b\Lambda r^2}
\ee
Substitution of this form into the equation leads to the following
conditions for $a, b$~:
\begin{eqnarray}
(d-3)a(a+b)+2(d-5)b&=&0 \label{2.3.2a}\\
3(d-3)a-(d-11)b+2(d-1)&=&0\: . \label{2.3.2b}
\end{eqnarray}
Exact solutions can then be obtained by solving a quadratic 
conditions in $a$ or $b$. Approximate
numerical values are presented below. For $d=4$ 
the old results \cite{BCC1,BCC2} are reproduced with
\be
\label{2.3.3}
a=\pm i1.732, \qquad b=-0.857 \mp i0.742\: .
\ee
Unlike in the Schwarzschild case, the $d=4$ 
dimensional solution of \re{2.18} is complex.

For the special case $d=5$ the eqations reduce to
\begin{eqnarray}
a(a+b)=0 \label{2.3.4a} \\
3(a+b)+4=0 \label{2.3.4b}
\end{eqnarray} 
whose only consistent solution,
\be
\label{2.3.5}
a=0, b=-(4/3)\: ,
\ee
leads to divergence in $K$ since one considers the domain $0\leq \Lambda r^2
\leq 1$.

>From $d=6$ onwards (exhibiting a behaviour complemetary to the Schwarzschild case) the
solutions become real. The solutions read, with $\epsilon = \pm 1$,
\be
\label{2.3.6}
a_{\epsilon} = {-(d-3)(d-4)+\epsilon\sqrt{(d-3)(5d-23)}\over
{(d-3)(d-5)}}
\ee
\be
\label{2.3.7}
b_{\epsilon} = {-(d^2-9d+26)+ 3\epsilon  \sqrt{(d-3)(5d-23)}
\over{(d-11) (d-5)}}
\ee

Only the solution corresponding to $\epsilon = -1$ leads to a
regular solution on $r\in [0,\bar \Lambda]
$ and for $5<d<11$. 
The single zero of the function $K(r)$ reads immediately
from the ansatz and the fact that $a_-$ is negative.

When solving numerically the equations for floating
values of $d$ for $d\rightarrow 5$ and $d\rightarrow 11$, we got 
evidences that the solution is running into problems.

The parameters $a,b$ of the regular solutions
 together with the value of the action
 (in fact of integral $I_{(d)}$  \re{2.2.9}
 now taken on $r \in [0,\Lambda^{-{1\over 2}}]$) 
 for the different $d$ are summarized in Table 1.

\begin{table}[h]
\caption{The values $a,b$ and the action for the
  solutions in the case p=1.}
\begin{center}
\begin{tabular}{|cccc|}
\hline
d &a&b&I\\
\hline
6  &-3.527 &4.349  &2.506\\
7  &-2.366 &4.098  &1.917\\
8  &-1.948 &5.073  &1.649\\
9  &-1.729 &7.558  &1.496\\
10 &-1.593 &15.448 &1.400\\
\hline
\end{tabular}
\end{center}
\end{table}

Again unlike in the Schwarzschild case, we did not find any solutions with two or more nodes, despite
our (numerical) efforts to do so. While this may signal the fact such solutions on deSitter backgrounds
do not exist, this is not necessarily case. In the latter case, it would be a challenge to find the multinode solutions.

\subsubsection{$p=2$ case}

In analogy with the $p=1$ case above, 
we would expect that there exist no real solutions to \re{2.18}
in $d=4p=8$. Similarly, we would expect that the solution 
in $d=4p+1=9$ would have divergent action/energy
integral $I_{(9)}$. Accordingly, we would expect 
to find (real) solutions for $d>4p+2=10$. This turns out
to be the case.
Surprisingly, the solution for $d=11$ turns out to 
be of the closed form \re{2.3.1}, with
\be
\label{2.3.9}
K(r) = {1-2\Lambda r^2\over{1+2\Lambda r^2}}\: .
\ee
This is in contrast to the $p=2$ Schwarzschild cases where all solutions were constructed numerically. All
the other solutions can be constructed only numerically.

We found numerical solutions with one node of the function 
$K(r)$ for $10 \leq d \leq 16$,
recovering the solution \re{2.3.8} numerically. 
The numerical approximation of their energy and of the
position $r_0$ of the node of $K(r)$ are given in Table 2.
We do not exhibit 
the profiles of the functions $K$
in these cases because they are very close to
the profile for the (analytically known) $p_1=1$ solution.

\begin{table}[h]
\caption{ The values of the node of the solutions and the action
  for the solutions in the case p=2}
\begin{center}
\begin{tabular}{|ccc|}
\hline
d &$\sqrt{\Lambda} r_0$&I \\
\hline
10 &0.61 &3.56\\
11 &$1/\sqrt{2}$ &2.67\\
12 &0.76&2.23\\
13 &0.79&1.96\\
14 &0.82&1.79\\
15 &0.84&1.66\\
16 &0.85&1.57\\
\hline
\end{tabular}
\end{center}
\end{table}

To summarize, we found one--node solutions on deSitter 
background in $4p+2\le d \le 8p+2$ for the $p=1$ YM
systems, and $4p+2\le d \le 8p$ for the $p=2$ YM systems.

\section{Selfdual YM on double--selfdual EH}

In Ref.~\cite{OT}, the result of Ref.~\cite{CD} for the $p=1$ case has been extended to the arbitrary $p$ case.
This states that in $4p$ dimensions, the $(p_1=)p$--th member of the YM hirearchy on the
double--selfdual background of the
$(p_2=)p$--th member of the EH hierarchy is selfdual. Unlike the non--selfdual solutions studied in the
previous Section which satisfied the YM equations \re{2.18a} on a fixed background, these solutions
satisfy the full gravitational--gauge field system taking into account the backreaction of the gravitational
system on the YM field~\cite{ft3}.

Our aim in this Section is to calculate the action densities of the $p=1$ and $p=2$ YM systems in
$d=4$ and $d=8$ respectively, for the purpose of comparing their actions with those of non--selfdual
solutions of these systems found in the previous Section.

The double-selfdual $2p$-form Riemann curvature in $d=4p$, whose metric automatically satisfies the
variational equations of the $p$--th EH system~\cite{CT1,OT,CT2}, yields the $2p$--form (single)
selfdual YM curvature in the chiral representation $so_{\pm}(4p)$ of $SO(4p)$ that satisfies the
variational equations of the $p$--th YM system. Following Refs.~\cite{CD,OT} this YM curvature is given by the Riemann tensor as follows 
\be
\label{3.1}
F_{\mu \nu}=-{1\over 2}\: R_{\mu \nu}^{ab} \: \Sigma_{ab}
\ee
where the Greek letters $\mu ,\nu$ are the coordinate indices and the early Latin letters $a,b$ the frame
indices, both running over $(1,2,3,...,4p)$. Renaming $4p$ as $0$, the coordinate index $\mu$ runs over $(i,0)$
and the frame index $a$ over $(m,0)$, where the Latin (coordinate) indices $i,j,k$ and (frame) indices
$m,n$ run over $1,2,..,4p-1$. $\Sigma_{ab}$ in \re{3.1} is defined by \re{2.1}.

While both Schwarzschild and deSitter metrics result in double--selfdual $2p$-form Riemann tensors, it is
only in the background of Schwarzschild that the YM fields have real non--selfdual solutions, which
were found in the perevious Section. Since selfdual YM fields on double--selfdual backgrounds are real,
it follows that for the purpose of comaring the latter with the former, only the Schwarzschild case is
relevant to the work of this Section. The EH systems supporting double--selfdual solutions in $4p$ dimensions
has been studied in Ref.~\cite{CT2}, which we exploit here.

Using the the Kerr-Schild parametrisation of the last Section, but now following the convention of
Ref.~\cite{CT2} with $C$ replaced by $2C$ and $L=l_0^2$, the
components of the Riemann tensor can be readily calculated
\begin{eqnarray}
R_{ij}^{mn} &=& \frac{2CL}{r^2}\delta_{[i}^m \delta_{j]}^n
-{C\over r}[(L' -{2L\over r}) +CLL']\hat x_{[i}\: \delta_{j]}^{[m}\hat
x^{n]}
\label{3.2a} \\
R_{i0}^{m0} &=& (1-CL){CL'\over r}\delta_i^m +C[(L''-{L'\over r})+
{CLL'\over r}]\hat x_i \hat x^m \label{3.2b} \\
R_{ij}^{m0} &=& \frac{C^2 LL'}{r}\hat x_{[i}\delta_{j]}^m \label{3.2c} \\
R_{i0}^{mn} &=& \frac{C^2 LL'}{r} \hat x^{[n}\delta_i^{m]} \label{3.2d}.
\end{eqnarray}

Substituting \re{3.2a}-\re{3.2d} into \re{3.1} we find the components of $F_{\mu \nu}$. Further using the
metric \re{2.4} we find $F^{\mu \nu}$. We list here only those components of the (covariant and contravariant)
curvature(s) that we will need below
\begin{eqnarray}
F^{ij} &=& -\frac{2CL}{r^2}\Sigma_{ij}
+{C\over r}\left[(1-CL)L'-{2L\over r}\right] \hat x_{[i}\Sigma_{j]k}\hat x_k
-\frac{C^2 LL'}{r}\hat x_{[i}\Sigma_{j]0} \label{3.3a} \\
F_{ij} &=& -\frac{2CL}{r^2}\Sigma_{ij} + {C\over r}\left[(1+CL)L'
-{2L\over r}\right]\hat x_{[i}\Sigma_{j]k}\hat x_k
-\frac{C^2 LL'}{r}\hat x_{[i}\Sigma_{j]0} \label{3.3b} \\
F_{k0} &=& -\frac{C^2 LL'}{r}\Sigma_{kl}\hat x_l -{C\over r}(1-CL)L'\Sigma_{k0}
-C\left[L''-(1-CL)\frac{L'}{r}\right]\hat x_k \hat x_l \Sigma_{l0} \label{3.3c}\ .
\end{eqnarray}
In addition to \re{3.3a} \re{3.3b} and \re{3.3c}, we will need the components $F^{lijk}$, $F_{lijk}$
and $F_{mnr0}$ of the $4$-form YM curvatures for the $p=2$ case, which can readily be calculated using
\re{2.16}. These are needed for the calculation of the action density.

Before giving the required action densities, we briefly verify that the fields given in \re{3.3a}-\re{3.3c}
actally lead to selfdual YM $2p$--forms. The two selfduality equations can be stated~\cite{T} as
\begin{eqnarray}
F^{ij}&=&\vep^{ijk}F_{k0} \label{3.4a} \\
F^{lijk}&=&\frac{1}{3!}\vep^{lijkmnr}F_{mnr0}\ . \label{3.4b}
\end{eqnarray}
Using the tensor--spinor identities
\begin{eqnarray}
\Sigma_{ij}&=&\vep_{ijk}\Sigma_{k0} \label{3.5a} \\
\Sigma_{lijk}&=&\frac{1}{3!}\vep_{lijkmnr}\Sigma_{mnr0} \label{3.5b}
\end{eqnarray}
in each case respectively, and \re{3.3a}-\re{3.3c}, we find the following two simple differential equations
and their solutions
\begin{eqnarray}
L''&=&\frac{2L}{r^2}\: \:  ,\: \: i.e. \: \qquad L=\frac{1}{r} \label{3.6a} \\
(L^2)''&=&\frac{12(L^2)}{r^2}      \: \:  ,\: \: i.e. \: \qquad L=\frac{1}{r^{\frac{3}{2}}} \label{3.6b}
\end{eqnarray}
for $p=1$ and  $p=2$ respectively. These agree, through $L=l_0^2$, with \re{2.8} as expected.

It follows from the selfduality equations \re{3.4a}-\re{3.4b} that the action densities defined in \re{p=1}
and \re{p=2} for the $p=1$ and $p=2$ cases simplify to
\begin{eqnarray}
\tilde {\cal S}_{(1)}&=&2\times \mbox{Tr} \: F^{ij}F_{ij} \label{3.7a} \\
\tilde {\cal S}_{(2)}&=&2\times \mbox{Tr} \: F^{lijk}F_{lijk} \label{3.7b}
\end{eqnarray}
in which the contributions of the terms $2\: \mbox{Tr} \: F^{k0}F_{k0}$ and $2\: \mbox{Tr} \: F^{mnr0}F_{mnr0}$
are absorbed in the factors $2$ on the right hand sides of \re{3.7a} and \re{3.7b}. Thus for this purpose, we
need only calculate the components $F^{ij}$, $F_{ij}$, $F^{ijkl}$ and $F_{ijkl}$ of the YM
curvature, and not $F^{ijk0}$ and $F_{ijk0}$. In the definitions \re{3.7a} and \re{3.7b} we have omitted factors
which would cancel with the angular volumes and the periods of the (Euclidean) time \re{i1}, which are necessary
to render the C-P charges of the static spherically symmetric self--dual solutions equal to one.

A direct calculation then yields
\begin{eqnarray}
\tilde {\cal S}_{(1)}&=& 2\times 2^{\frac{d-6}{2}}
(d-2)\frac{C^2}{r^2}[2L'^2 +4(d-3)\frac{L^2}{r^2}] \label{en1-2} \label{3.8a} \\
\tilde {\cal S}_{(2)}&=& 2\times 2^{\frac{d-10}{2}}
(d-2)(d-3)(d-4)\frac{4^2 C^4}{r^6}L^2 [L'^2 +(d-5)\frac{L^2}{r^2}] \label{3.8b}\: ,
\end{eqnarray}
where it is understood that $d=4$ in \re{3.8a} and $d=8$ in \re{3.8b}, and we have
left $d$ in these expression by way of highlighting their relations to \re{2.19}
and \re{2.20}.

Substituting the solutions \re{3.6a} and \re{3.6b} in \re{3.8a} and \re{3.8b},
performing the angular integrations and the (Euclidean) time integrations over one
period \re{i1}, there remains only to perform the radial integrations we to evaluate
the action integrals
\be
\label{3.9}
\tilde S_{(p)}=\frac{P(p_2)\Omega_{(d-2)}}{c_{(p)}}\: \int_1^{\infty} \tilde {\cal S}_{(p)}\: 
r^{4p-2} \dd r \ .
\ee
In \re{3.9}, $c_{(p)}$ is a normalisation constant. The the angular
integration over the $d-2$ dimensional angular volume gives
\be
\label{3.10}
\Omega_{(d-2)} = \frac{2 \pi^{(d-1)/2}}{\Gamma((d-1)/2)}
\ee
The factor $P_{(p_2)}$, namely the period \re{i1}, is conributed by the (Euclidean)
time integration. $c_{(p)}$ is to be chosen
such that the Chern-Pontryagin (C-P) charge of the selfdual
spherically symmetric field configurations be normalised to unity. We calculate this
normalisation factor for the $p_1=p_2=1$ and $p_1=p_2=2$ cases in $d=4$ and $d=8$
dimensions respectively. The scale factor $C$ in \re{3.8a} and \re{3.8b} is fixed to
$C=1/2$, as was done in the evaluation of the components of the Riemann tensor in
\re{3.2a}-\re{3.2d}.

Substituting \re{3.6a} for $L(r)$ in \re{3.8a}, with $d=4$, and performing the radial
integral, we find
\be
\label{3.11}
c_{(1)}=P_{(1)}\Omega_{(2)} \ .
\ee
Similarly performing the radial integral of \re{3.8b} with $L(r)$ given by \re{3.6b} and $d=8$, we find
\be
\label{3.12}
c_{(2)}=90\ P_{(2)}\Omega_{(6)} \ .
\ee

We now compare the values of the actions of the non--selfdual solutions by performing the four and
eight dimensional integrals of \re{2.19} and \re{2.20} using the normalisations \re{3.11} and \re{3.12},
respectively. The results of the actual integrations, for both one--node and two--node solutions respectively,
are listed in \re{2.2.10a} and \re{2.2.12b} for the $p_1=1$ and $p-1=2$ cases in that order. The resulting
action integrals, analogous to \re{3.9}
\be
\label{3.13}
S_{(p)}=\frac{P(p_2)\Omega_{(d-2)}}{c_{(p)}}\: \int_1^{\infty} {\cal S}_{(p)}\:  r^{4p-2} \dd r \ ,
\ee
with ${\cal S}_{(p)}$ given by \re{2.19},\re{2.20},  \re{2.2.9},\re{2.2.10},  \re{2.2.10a},\re{2.2.12b}
take the values and \re{3.11},\re{3.12}
\begin{eqnarray}
S_{(1)}^{(1)}= 0.959   \qquad   & S_{(1)}^{(2)}=0.992 \label{3.14a} \\
S_{(2)}^{(1)}= 1.587   \qquad   & S_{(2)}^{(2)}=1.933  \ . 
\label{3.14b}
\end{eqnarray}
The superscripts in \re{3.14a}-\re{3.14b} pertain to the number of nodes, as in \re{2.2.10a},\re{2.2.12b}.



The magnitudes of these actions are to be compared relative to the unit valued actions of
their selfdual partners. In the latter case, these are simply equal to the unit C-P charges, while in the
non--selfdual cases the C-P charges equal $0$ by virtue of the vanishing of $F_{i0}$, \re{2.12a}. The only
quantitative conclusion that can be made is, that it appears the actions grow with increasing number of nodes,
which is not surprising.

\section{Summary and discussion}

Since the known real solutions to the $4p$-dimensional $p$-YM hierarchy are self--dual, it is of some interest
to construct non--selfdual solutions, and this is what we have done by considering these systems on fixed
curved backgrounds. YM fields on fixed curved backgrounds can be regarded as approximations to the fully
interacting YM--gravitational fields, but in this paper we have restricted to the fixed gravitational
backgrounds. Nonetheless we have discovered properties that are akin to those of the usual $4$-dimensional
Einstein--YM fields studied by Bartnik and McKinnon~\cite{BM}. Specifically, we have found that there
exist solutions for which the radial function parametrising the YM field has one-- and two--nodes. It is
quite likely that there should be a sequence of solutions, like in \cite{BM}, with increasing number of
nodes, but we did not search for these.

We have presented {non--selfdual} solutions to the $p_1$-hierarchy of Yang--Mills (YM) systems on the
fixed backgrounds of Schwarzchild and deSitter spaces. The specific constructions are made for the 
$p_1=1$ and the $p_1=2$ systems in various dimensions, but the qualitative conclusions arrived at are
expected to remain valid for arbitrary $p_1$.

The fixed Schwarzschild and deSitter metrics employed are
the solutions of the vacuum gravitational equations of the $p_2$-hierarchy of Einstein--Hilbert (EH) systems.
Thus, for each of the two cases $p_1 =1,2$, we have employed $p_2 =1,2,3$ fixed backgrounds consistent with
the requirement of asymptotic flatness of the metric. Since the deSitter metrics for all $p_2$-EH systems
are identical, this multiple choice of curved background is relevant only for the Schwarzschild metrics
\re{2.8}, given in \cite{CT2}. But in that case (i.e. Schwarzschild) the non--selfdual solutions to the $p_1$-YM
systems we construct are real only in dimensions $d=4p_1$ whence it follows that for the $p_1 =1$ case in
$4$-dimensions, only the $p_2 =1$ background can be used consistently with the asymptotic flatness condition.
For the $p_1 =2$ Schwarzschild case, we use all possible backgrounds $p_2 =1,2,3$. An interesting observation that
can be made is, that for a given $p_1$-YM system on a $p_2$-EH background, the energy decreases with increasing
$p_2$ of the background. This was found to be true for both one--node and two--node solutions, {\it cf.}
\re{2.2.12a}-\re{2.2.12c}.

In the $p_1=1$ Schwarzschild case, the one--node solutions are constructed both in closed form and numerically,
while the two--node solutions of this system, as well as all (one-- and two--node) solutions in the $p=2$
case were possible to construct only numerically. The profiles of the functions $K(r)$ are exhibited in Figs. 1,2.

In the $p_1=1$ deSitter case, the one--node solutions are constructed both in closed form and numerically, while
the solutions in the $p_1=2$ deSitter case these are 
constructed only numerically. No  solutions with more than one mode were
found in these cases numerically, but we do not 
know if this indicates the non--existence of such solutions. If
they exist, then it would be a challenge to find these, but it is possible that they do not since the two
backgrounds, Schwarzschild on which two--node solutions 
exist and deSitter, are qualitatively quite different
from each other. Notably the intervals on which they are defined are respectively non--compact and compact. In
the deSitter cases we did not exhibit the profiles of the functions $K(r)$ for these solutions as these all have
one--node only, and, they are very close to the closed form solution of the $p_1=1$ case. Instead we have
listed their properties, namely the position of the (single) 
node and the energy integral, in Tables 1 and 2, for
the $p_1=1$ and $p_1=2$ cases respectively.

That the solutions discussed above are non--selfdual is not a matter of note until one considers that in
the specific dimensions $4p_1=4p$ the $p_1=p$-YM fields on double--selfdual ($2p$-form Riemann tensor)
$p_2=p$-EH background, are (single) selfdual. In these cases where $p_1=p_2=p$, the action of the gravitational
system vanishes due to the vanishing of the stress tensor of the selfdual YM fields, so that the total action
equals the Chern--Pontryagin (C-P) charge. In the spherically symmetric Schwarzschild cases considered,
this is the unit C-P charge. The comparison of this action with the action pertaining to the
corresponding non--selfdual solution reveals an interesting feature. (Note that the non--selfdual solutions
have zero C-P charge.) It was found that for the 4-dimensional ($p_1 =1$) YM system on $p_2=1$
Schwarzschild background, the actions of the nonselfdual solutions with one and two nodes respectively
were equal to $0.959$ and $0992$, which are slightly less than the unit selfdual action. (Of these
the exact value of the one--node action was found in Ref.~\cite{BCC1}.) We have found that for the $p_1=2$ YM
system on $p_2=2$ Schwarzschild background in 8 dimensions,
the actions of the one node and two node solutions respectively
are equal to $1.587$ and $1.933$, which are appreciably larger than the unit magnitudes of the corresponding
selfdual actions.


In addition, we have exhibited anti-deSitter solutions of the $p$-YM 
systems in $4p$ dimensions which can
be related with meron-type solutions in flat space-time 
through conformal transformations. 
Very recently, soliton-types of solutions
have been obtained for the Einstein-Yang-Mills equations
with asymptotically anti-de Sitter space \cite{hosotami}

There remain two obvious directions in which the study of the present work can be developed to complete
and enhance the conclusions drawn here. The most obvious extension is to find the (non--selfdual) solutions
of the full YM-EH systems taking into account fully the backreaction of gravity on the gauge fields,
rather than studying the gauge fields on a fixed background. The actions of the solutions in that case would
differ from what we have found, but maybe not very much. What is important is that the topological (C-P) charges
of these solutions will still be equal to zero so that like those on the fixed backgrounds, these solutions
will also describe (unstable) saddle points. Another direction, is to study the solutions of combined $p_2$ EH
systems in dimensions $d>4p_2$. (Note that in $d=4p_2$ the 
$p_2$-EH system \re{iip=1}-\re{iip=3} reduces to the
Euler-Hirzebruch topological density which has no dynamics.) It would be interesting to find out how the inclusion
of a higher $p_2$-EH term (presumably with a small coupling constant) would modify the metric and other properties
of the leading $p_2$ system. In addition, one can contemplate the addition to these types of systems, members of
the $p_1$-YM hierarchy.

\medskip

\noindent
{\bf Acknowledgements}
We are grateful to B. Kleihaus for showing us the solution \re{2.3.9}. This work was carried out in the framework
of project IC/99/070 of Enterprise--Ireland.

\newpage
\section*{Appendix: A nonselfdual solution for $AdS_4$ background.}
\setcounter{equation}{0}
\renewcommand{\theequation}{A.\arabic{equation}}

For $d=4$ a divergent solution in deSitter background was shown \cite{BCC1,C} to be related through
conformal transformations to meron-type solutions in flat space-time. This, particularly
simple, solution corresponds (with $\Lambda >0$) to
\begin{equation}
\label{a1}
N=(1-\Lambda r^2), \quad K=N^{-\frac{1}{2}}=(1-\Lambda r^2)^{-\frac{1}{2}}
\end{equation}
Here we note that changing the sign before $\Lambda$ one obtains the anti-deSitter case
($AdS_4$)  and
\begin{equation}
\label{a2}
N=(1+\Lambda r^2), \quad K=N^{-\frac{1}{2}}=(1+\Lambda r^2)^{-\frac{1}{2}}\ .
\end{equation}
This provides a solution of \re{2.18} with $p=1$ and $d=4$, namely of
\begin{equation}
\label{a3}
(NK')'=r^{-2}K(K^2 -1)
\end{equation}
where now $K$ is no longer singular.

Remarkably, we note that \re{a2} is a solution to
the $p$--YM system in $4p$ dimensions, satisfying the equation
\be
\label{a4}
[N(K^2 -1)^{p-1}K']'=(2p-1)r^{-2}K(K^2 -1)^p
\ee
that results from \re{2.18} setting $d=4p$ in it. This is not surprising since it is known that the $p$-YM
system on $\R^{4p}$ supports meron solutions~\cite{OT2}.

One can now evaluate the total actions for these
$4p$-dimensional solutions \re{a2} and study their different properties. We will restrict this to the ($p=1$)
$4$-dimensional case, setting $\Lambda =1$ henceforth in this Appendix for simplicity.

The radial integral of \re{2.19}, in the absence of the horizon, should
now be replaced by
\begin{equation}
\tilde{I}_{(4)}=\int_0^{\infty} dx x^{-2}\Bigl(2N\Bigl(x\frac{dK}{dx}\Bigr)^2 + \Bigl(K^2
-1\Bigr)^2 \Bigr)
\end{equation}
where
$$N=(1+x^2), \quad K=(1+x^2)^{-\frac{1}{2}}$$
One obtains
\begin{equation}
\tilde{I}_{(4)}= 3\int _0^{\infty} \frac{x^2}{(1+x^2)^2}dx =\frac{3\pi}{4}
\end{equation}
Thus, corresponding to \re{2.19}, one obtains a finite spatial integral
 \begin{equation}
\frac{3\pi^2}{2}
\end{equation}
The factor from the time integration depends on the chosen context. Now there is no horizon
to be desingularized and the discussion of Sec.1 is not directly relevant. But one can
start by considering the hypersurface
\begin{equation}
-t_1^2-t_2^2+x_1^2+x_2^2+x_3^2=-1
\end{equation}
In terms of the spherical coordinates
\begin{eqnarray}
(x_1,x_2,x_3)&&\rightarrow (r,\theta,\phi){\nonumber}\\
(t_1,t_2)&&\rightarrow (T,\psi)
\end{eqnarray}
the metric on the hypersurface
$$r^2 - T^2 =-1$$
is
\begin{equation}
ds^2 = -(1+r^2)d\psi^2 + (1+r^2)^{-1}dr^2+ r^2 d\Omega_2
\end{equation}
In this context the $\psi$-integration gives a factor $2\pi$ and one obtains a total action
\begin{equation}
3\pi^3
\end{equation}
But often it is preferable to consider the covering space $(CAdS)$ replacing $\psi \in
S^1$ by $t\in R$. Then the action is evidently divergent.

\newpage
\centerline{Figure Captions}
\begin{itemize}

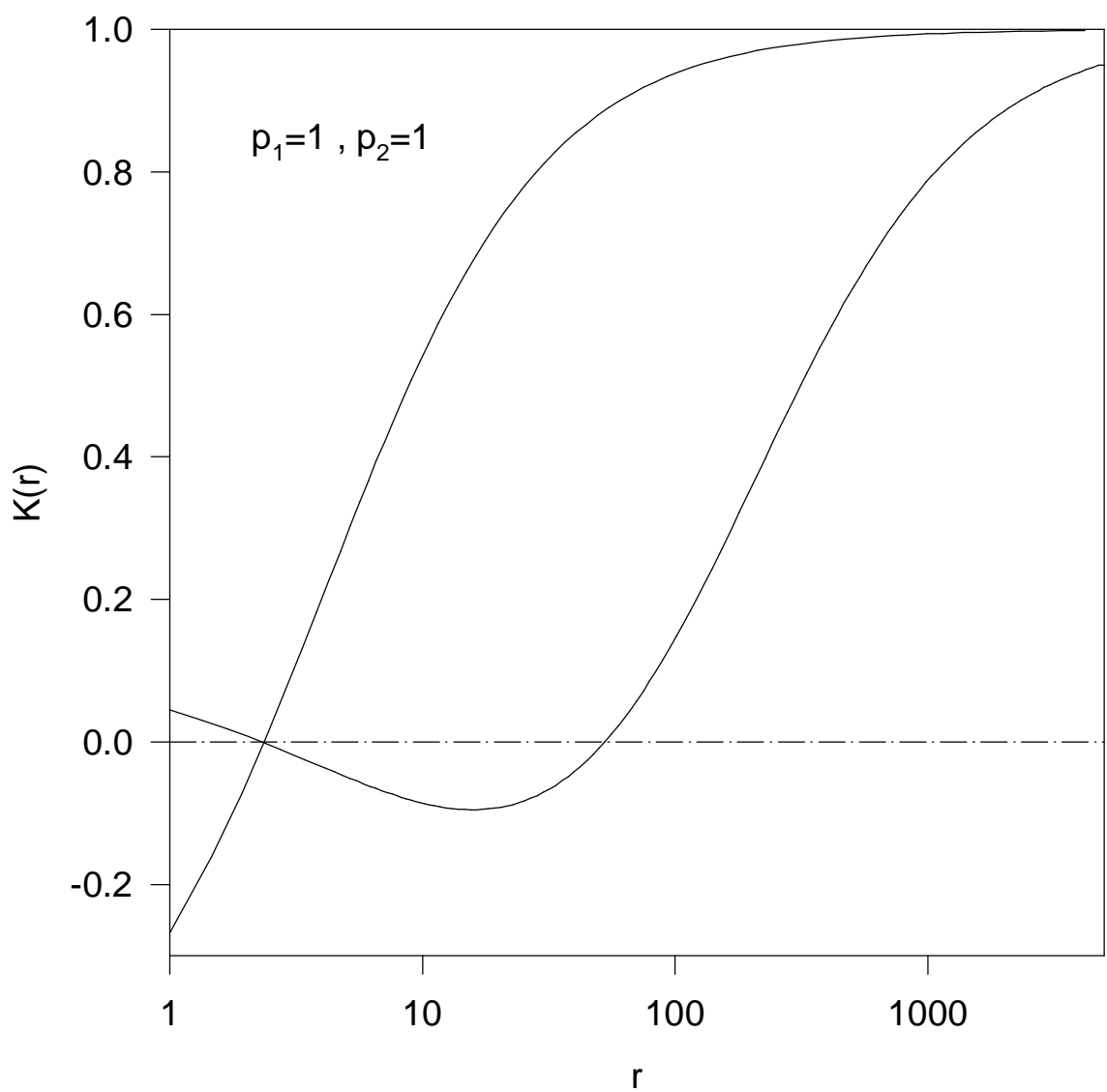
\item [Figure 1] 
The profiles of two solutions of Eq. (\ref{2.18a}) 
(one with one node and
one with two nodes) for the case $p_1=1,p_2=1$.

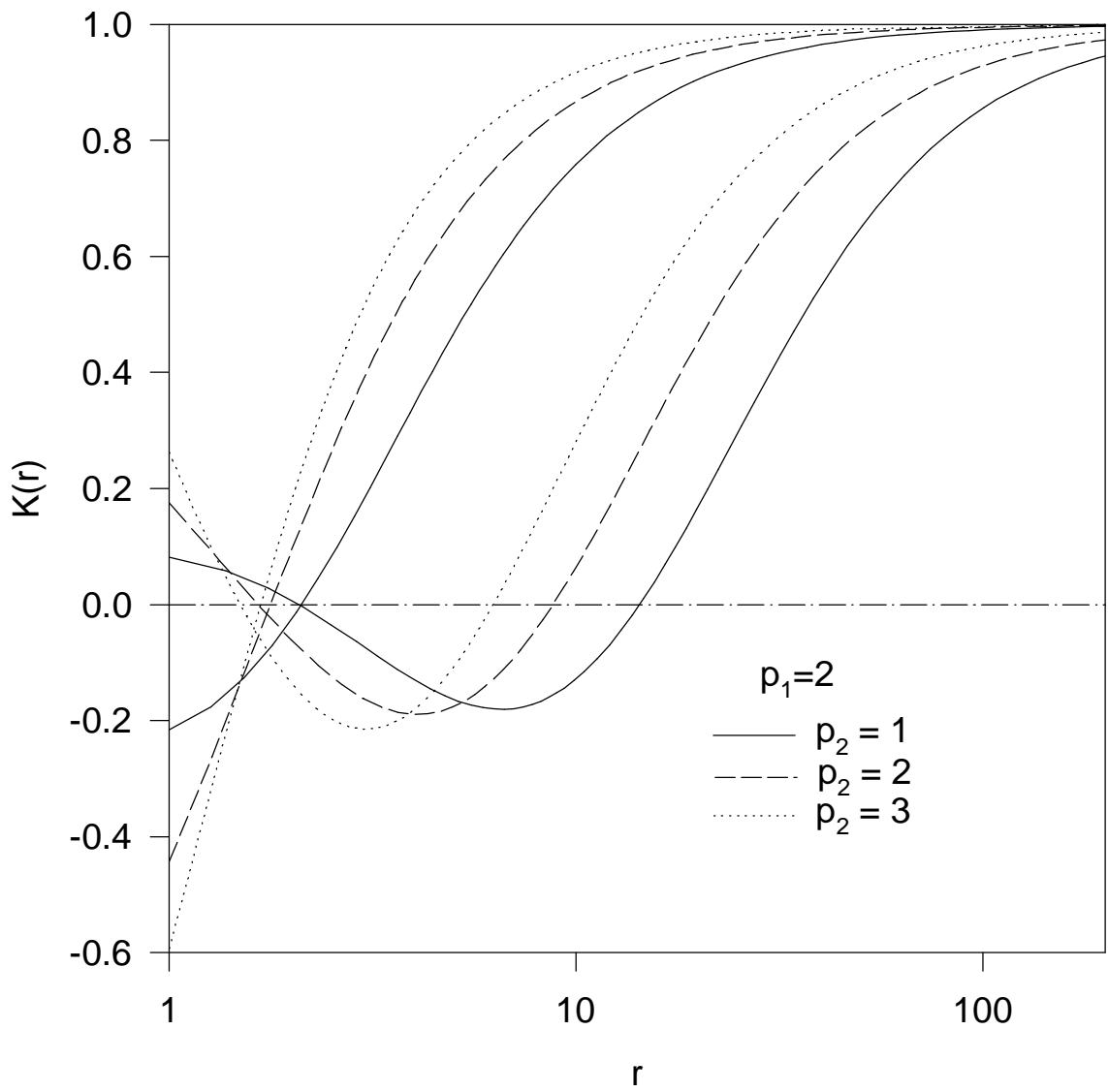
\item [Figure 2]
The profiles of two solutions of Eq. (\ref{2.18a}) 
(one with one node and
one with two nodes) for $p_1=2$ and for
$p_2=1$ (solid), $p_2=2$ (dashed), $p_2=3$ (dotted).
\end{itemize}
\clearpage
\newpage




\begin{thebibliography}{99}
\bibitem{ft1}
Note that both Schwarzschild and
deSitter Riemann curvatures are known to be double--selfdual in $4$ dimensions and hence there
exist selfdual YM solutions~\cite{CD}
\bibitem{ft2}
Anticipating the notation to be adopted below in \re{2.8}, \re{2.19}
and \re{2.11}, ${\cal K}$ can be read off $N=1-({\cal K}/r)^{\frac{d-2p_2 -1}{p_2}}$.
\bibitem{ft3}
As it happens in this case, this backreaction is nil since the
stress energy--momentum tensor vanishes by virtue of the YM selfduality.
\end{thebibliography}
\end{document}